\documentclass[journal,onecolumn,11pt]{IEEEtran}

\usepackage{amsmath,amssymb,amsthm,mathtools}
\usepackage{microtype}
\usepackage{cite}
\usepackage{float}
\usepackage{tikz}
\usepackage[colorlinks=true,citecolor=blue,linkcolor=black,urlcolor=black]{hyperref}

\allowdisplaybreaks
\newcommand{\F}{\mathbb{F}}
\newcommand{\LCS}{\operatorname{LCS}}
\newcommand{\RS}{\operatorname{RS}}
\newcommand{\eps}{\varepsilon}
\newcommand{\cC}{\mathcal{C}}
\newcommand{\one}{\mathbf{1}}

\theoremstyle{plain}
\newtheorem{theorem}{Theorem}
\newtheorem{lemma}[theorem]{Lemma}
\newtheorem{proposition}[theorem]{Proposition}
\newtheorem{corollary}[theorem]{Corollary}
\theoremstyle{definition}
\newtheorem{example}[theorem]{Example}
\newtheorem{remark}[theorem]{Remark}

\title{The Insertion-Deletion Distance of Linear Codes under Random Coordinate Permutations}
\author{Hengfeng~Liu, Chunming~Tang, and Zhengchun~Zhou%

\thanks{Hengfeng Liu is with the School of Mathematics,
Southwest Jiaotong University, Chengdu, China
(e-mail: hengfengliu@163.com).}%
\thanks{Chunming Tang and Zhengchun Zhou are with the School of
Information Science and Technology, Southwest Jiaotong University,
Chengdu, China
(e-mail: tangchunmingmath@163.com; zzc@swjtu.edu.cn).}}

\begin{document}
\maketitle

\begin{abstract}
We study the maximum insertion-deletion (insdel) distance attainable by permuting the coordinates of a fixed $[n,k]_q$ linear code. Exploiting the connection between common subsequences and noncrossing matchings, we establish probabilistic lower bounds under random coordinate permutations over large alphabets. In particular, if the maximum symbol multiplicity among
nonconstant codewords is $O(n/q)$, a uniformly random
coordinate permutation yields insdel distance at least
$2n-O(n/\sqrt q)$ with high probability for sufficiently large $q$. Furthermore, for projective linear codes satisfying the same
symbol multiplicity condition, we show that the maximum
attainable insdel distance is $2n-\Theta(n/\sqrt q)$. As an application, for $[n,k]_q$ MDS codes containing the
all-one vector, random coordinate permutations yield insdel
distance at least $2n-O(\sqrt{kn})$ with high probability
whenever $(k-1)(\log q)^2=o(n)$.
In particular, for full-length Reed--Solomon (RS) codes,
we obtain the lower bound $2q-O(\sqrt{kq})$ with high probability
for $k=o(q/(\log q)^2)$, improving previously known lower bounds
while allowing a wider range of dimensions.
Finally, we give explicit constructions of full-length RS codes
over square fields, attaining the order-optimal insdel distance
$2q-\Theta(\sqrt q)$.

\end{abstract}


\section{Introduction}

Synchronization errors in communication and storage systems may cause symbols to be inserted into or deleted from a transmitted sequence. The received sequence is then no longer synchronized with the transmitted one, and these errors are referred to as insertion and deletion errors (insdel errors, for short).  In recent years, insdel codes have received considerable attention due to the occurrence
of insdel errors in high-density storage systems and wireless communication systems, such as DNA storage, language processing,
and racetrack memory error correction \cite{CheraghchiRibeiro,HaeuplerShahrasbiSurvey}.

The theory of insdel codes starts with the seminal work of Levenshtein
\cite{Levenshtein1966}, who proved that the Varshamov--Tenengolts (VT)
codes \cite{VarshamovTenengolts1965} can correct a single insertion or
deletion.
For a fixed number of synchronization errors, a challenging problem is
to construct explicit codes with redundancy close to the minimum possible.
Considerable progress has been made for correcting a fixed number of
deletions, particularly in the two-deletion case
\cite{BrakensiekGuruswamiZbarsky2018,SimaBruck2021,GuruswamiHastad2021,LiuTjuawinataXing2024}.
Nevertheless, the optimal redundancy for correcting a fixed number of
deletions remains unknown in general. Although many of the known insdel codes are nonlinear, linear
codes are of particular interest due to their compact representations and
efficient encoding.  Their insdel error-correcting capability, however, is subject to stronger
restrictions. It was proved in  \cite{AbdelGhaffarFerreiraCheng}  that a linear code
capable of correcting even a single deletion must have rate at most
$1/2$. 

For vectors $\mathbf u,\mathbf v\in\F_q^n$, the Hamming distance $d_{\rm H}(\mathbf u,\mathbf v)$ counts the coordinates in which they differ. Their insdel distance $d_{\rm I}(\mathbf u,\mathbf v)$ is the minimum number of insertions and deletions required to transform one into the other.  For a fixed linear code,  permutation of coordinates leads to equivalent codes under the Hamming metric, but they may have different  insdel distances. In fact, the insdel distance can take several distinct values even
for a short Reed--Solomon code. For example, over $\F_{13}$, take the
evaluation vector $\boldsymbol\alpha=(0,1,\ldots,12)$ and consider the three-dimensional full-length Reed--Solomon code $\RS_{13,3}(\boldsymbol\alpha
=\bigl\{(f(\alpha_i))_{i=1}^{13}:
f\in\F_{13}[x],\ \deg f<3\bigr\}$.
This code  has Hamming distance $11$, but its insdel distances
over the $13!=6\,227\,020\,800$ permutations of the coordinates
are distributed as shown in Table~\ref{tab:ordering-distribution}.

\begin{table}[t]
	\centering
	\caption{Insdel distance distribution over all coordinate permutations
		of $\RS_{13,3}(\boldsymbol\alpha)$ over $\F_{13}$, where
		$\boldsymbol\alpha=(0,1,\ldots,12)$. }
	\label{tab:ordering-distribution}
	\small
	\setlength{\tabcolsep}{6pt}
	\renewcommand{\arraystretch}{1.2}
	\begin{tabular}{|c|r|r|}
		\hline
		\makebox[1.6cm][c]{$d_{\rm I}$} &
		Permutations & Percentage (\%)\\
		\hline
		$2$  & $382\,668$           & $0.00614528$\\
		\hline
		$4$  & $42\,183\,492$       & $0.67742655$\\
		\hline
		$6$  & $1\,422\,576\,324$   & $22.84521555$\\
		\hline
		$8$  & $4\,728\,796\,020$   & $75.93994258$\\
		\hline
		$10$ & $33\,082\,296$       & $0.53127004$\\
		\hline
	\end{tabular}
\end{table}

Several upper bounds on the insdel distance of linear codes have been established in recent years. For an $[n,k]_q$ linear code $\cC$, Cheng et
al.~\cite{ChengGuruswamiHaeuplerLi} proposed the half-Singleton bound, and Chen \cite{Chen2022} gave its non-asymptotic version
$d_{\rm I}(\cC)\le\max\{2(n-2k+2),2\}$. Later, Ji et
al.~\cite{JiZhengChenWang2023} sharpened it to
$d_{\rm I}(\cC)\le\max\{2(n-2k+1),2\}$ when $\one\notin\cC$, which is called the strict half-Singleton bound.
More upper bounds that hold for any coordinate ordering were obtained in \cite{ChenZhang2022,Chen2022,XieChenQuLiu2024}. In addition to establishing upper bounds, Chen~\cite{Chen2022} also highlighted the challenge of obtaining good lower bounds on the insdel distances of well-constructed binary linear codes. In this direction, Qiu et al.~\cite{QiuZhuLiSun2026} established lower bounds for certain subcodes of binary first-order Reed--Muller codes and simplex codes.

Although determining the insdel distance of a general linear code is challenging, substantial progress has been made for Reed--Solomon (RS) codes.
Several bounds and constructions have been developed for RS and generalized RS codes \cite{DucLiuTjuawinataXing,ChenZhang2022,LiuTjuawinata}, with particular attention paid to attaining the half-Singleton bound. For general dimensions, the existence of RS codes attaining this bound over sufficiently large fields was established, together with deterministic constructions  \cite{ConShpilkaTamo2023,Liu2024}. For dimension two, the minimum field size required to attain the bound was determined to be $\Theta(n^3)$ \cite{ConShpilkaTamo2023,ConShpilkaTamo2024}. More recently, Con et al.~\cite{ConGuoLiZhang2025} showed that random RS codes approach the half-Singleton bound with high probability over linear-sized alphabets.

Full-length RS codes provide a natural setting for studying the effect of coordinate permutations on insdel distance. Their evaluation set is  $\F_q$, so the choice of a code $\RS_{q,k}(\boldsymbol\alpha)$ amounts to ordering these evaluation points. Very recently, Beelen et al.~\cite{BeelenEtAl} initiated this study. For general $k\ge2$, they proved that if $q\ge10ek^3e^{6k}$, there exists an ordering $\boldsymbol\alpha$ of $\F_q$ such that the corresponding full-length RS code $\RS_{q,k}(\boldsymbol\alpha)$ has insdel distance at least $q/(5k)$.
For $k=2$, they showed that for sufficiently large $q$, a uniformly random ordering yields insdel distance at least $2(1-\delta)q$ with high probability for every fixed $\delta\in(0,1)$. However, finding an explicit ordering with comparably large insdel distance remains challenging even for two-dimensional full-length RS codes, as noted in the open problem in \cite{BeelenEtAl}. The corresponding question for general linear codes, without the benefit of the special structure of RS codes, is more challenging still.

We denote the maximum insdel distance $d_{\rm I}^{\max}(\cC)$ attainable by permuting the coordinates of a fixed linear code over  large alphabets. In this paper, we address the following two natural questions:

\vspace{0.5em}
\noindent\emph{Question 1.}
For a fixed linear code $\cC$, how large can
$d_{\rm I}^{\max}(\cC)$ be, and what insdel distance
can be achieved by a random coordinate permutation?

\vspace{0.5em}
\noindent\emph{Question 2.}
When random coordinate permutations yield large insdel
distance with high probability, can we explicitly construct
orderings that achieve comparable distances?
\vspace{0.5em}

Our main results are as follows.
\begin{itemize}
   \item \textbf{Random coordinate permutations yield large insdel distance with high probability.}  We establish probabilistic lower bounds on the insdel distance
    of linear codes under random coordinate permutations in terms
    of their minimum Hamming distance and maximum symbol multiplicity,
    with explicit exponential estimates for the failure probability
    (Theorem~\ref{thm:linear-random}).
    For sufficiently large $q$, a uniformly random coordinate
    permutation yields insdel distance at least $2n-O(n/\sqrt q)$
    with high probability when the maximum symbol multiplicity
    among nonconstant codewords is $O(n/q)$
    (Corollary~\ref{cor:linear-square-root}).
    Our proof uses the representation of common subsequences
    between codewords as noncrossing matchings in their relation graphs.
    As an application, for MDS codes containing the all-one vector, we obtain the lower bound
    $2n-2(k-1)-(2e+o(1))\sqrt{(k-1)n}$ with high probability whenever $(k-1)(\log q)^2=o(n)$.
    In particular, for full-length RS codes, taking $n=q$
    gives insdel distance at least $2q-O(\sqrt{kq})$ with high
    probability whenever $k=o(q/(\log q)^2)$.
    For $k\ge3$, this improves the lower bound $q/(5k)$ established by
    Beelen et al.~\cite[Thm.~2]{BeelenEtAl} under
    $q\ge10ek^3e^{6k}$ and allows a wider range of dimensions.
    For $k=2$, the bound with leading constant $2e$ also follows
    from a refinement of their counting argument
    (Remark~\ref{rem:rs-beelen}).

 \item \textbf{Maximum insdel distance of order \(2n-\Theta(n/\sqrt q)\).} We prove an alphabet-dependent upper bound for projective linear codes
   that is independent of coordinate ordering.
   When the maximum symbol multiplicity among nonconstant codewords is
   $O(n/q)$, this bound is asymptotically tighter than previous
   bounds compared in Table~\ref{tab:upper-comparison}:
   the previously known bounds have deficits of $O(n/q)$ from $2n$,
   whereas our bound has a deficit of $\Theta(n/\sqrt q)$.
   Together with our probabilistic lower bound, this determines
   $d_{\rm I}^{\max}(\cC)=2n-\Theta(n/\sqrt q)$
   under the same condition.

\item \textbf{Explicit orderings achieve insdel distance $2q-\Theta(\sqrt q)$ for full-length RS codes.} We construct deterministic orderings of full-length RS codes with distance $2q-\Theta(\sqrt q)$ over square fields.
    The construction groups evaluation points into polynomial level sets over the subfield and uses B\'ezout's theorem for algebraic curves to control common subsequences.
    It applies to every fixed dimension $k\ge2$ when the characteristic exceeds $k+1$.
    A modified construction covers dimension two in every characteristic.
    These orderings are computable in time polynomial in $q$ and answer the explicit-ordering question over square fields.
\end{itemize}

\subsection*{Organization of the paper}

Section~\ref{sec:prelim} presents the tools and auxiliary results
needed for the subsequent proofs. Section~\ref{sec:bounds} establishes
probabilistic lower bounds and applies them to MDS codes, full-length
RS codes, and two non-MDS evaluation codes, and we also prove an
upper bound for projective linear codes and combine it with the lower
bounds to determine the maximum attainable insdel distance under a
condition on symbol multiplicities. Section~\ref{sec:explicit} gives
explicit orderings for full-length RS codes over square fields.
We conclude with open questions in Section~\ref{sec:conclusion}.

\section{Preliminaries}\label{sec:prelim}
In this section, we recall basic facts about insdel distance,
the representation of common subsequences by noncrossing matchings,
and the results on function fields and plane curves needed
for our explicit constructions.
\subsection{ Insdel Distance of Linear Codes}\label{subsec:code-prelim}

A sequence $(a_1,\ldots,a_r)$ is a \emph{common subsequence} of
$\mathbf u,\mathbf v\in\F_q^n$ if there exist indices $1\le i_1<\cdots<i_r\le n$ and  $1\le j_1<\cdots<j_r\le n$ such that $u_{i_t}=v_{j_t}=a_t$ for every $1\le t\le r$.  Let $\LCS(\mathbf u,\mathbf v)$ denote the length of a longest common subsequence between them. The following well-known relation between insdel distance and LCS has been proved in \cite{HaeuplerShahrasbi2017}.
\begin{lemma}[\cite{HaeuplerShahrasbi2017}]\label{lem:distance-lcs}
	For any $\mathbf u,\mathbf v\in\F_q^n$,
	\begin{equation}\label{eq:distance-lcs}
		d_{\rm I}(\mathbf u,\mathbf v)
		=2\bigl(n-\LCS(\mathbf u,\mathbf v)\bigr).
	\end{equation}
	Consequently, for a code $\cC\subseteq\F_q^n$, we have
	\[
	d_{\rm I}(\cC)=2\bigl(n-\LCS(\cC)\bigr),
	\]
	where $	\LCS(\cC):=\displaystyle\max_{\substack{\mathbf u,\mathbf v\in\cC\\\mathbf u\ne\mathbf v}}
	\LCS(\mathbf u,\mathbf v)$.
\end{lemma}

For a  code of fixed length, correcting up to $t$ deletions is
equivalent to correcting up to $t$ insertions \cite{Levenshtein1966}.
The following criterion characterizes this correction capability
in terms of the longest common subsequence \cite{BeelenEtAl}, which follows directly from Lemma \ref{lem:distance-lcs}.

\begin{proposition}[\cite{BeelenEtAl}]
	\label{prop:insdel-correction}
	Let $\cC\subseteq\F_q^n$ be a code with at least two codewords,
	and let $t$ be a nonnegative integer. Then $\cC$ can 
	correct   $t$ insertions and deletions
	if and only if
	\[
	\LCS(\cC)\le n-t-1,
	\]
	or equivalently, $d_{\rm I}(\cC)\ge 2t+2$.
\end{proposition}

\subsection{Common Subsequences of Words and Noncrossing Matchings in Bipartite Graphs}\label{subsec:graph-prelim}
For $\mathbf u,\mathbf v\in\F_q^n$, a common subsequence of
length $r$ can be represented by a \emph{noncrossing matching} of
size $r$ in a bipartite graph \cite{KiwiLoeblMatousek2005}.
We first recall the relevant notions in matching, the reader is referred to \cite{LovaszPlummer} for more details.

All directed graphs considered below are finite.
For a directed graph $D=(V,E)$, a \emph{directed matching} is a set of
directed edges with pairwise distinct tails and pairwise distinct heads.
Let $V^-$ and $V^+$ be disjoint copies of $V$.
The \emph{bipartite representation} $\mathcal{B}(D)$ is the bipartite
graph with vertex classes $V^-$ and $V^+$, where $i^-$ and $j^+$
are adjacent exactly when $(i,j)\in E(D)$ \cite{LovaszPlummer}.
Conversely, every bipartite graph on these two copies determines
a directed graph on $V$. This correspondence gives a bijection between
matchings in $\mathcal{B}(D)$ and directed matchings in $D$.
The diagonal edges $i^-i^+$ correspond to loops $(i,i)$ in $D$,
so deleting all diagonal edges corresponds to removing all loops.
Throughout, we write $(i,j)$ for both the edge $i^-j^+$ and
its corresponding directed edge in $D$.

Fix a total order $\prec$ on $V$, which induces the same
order on $V^-$ and $V^+$. A matching
$M=\{(i_s,j_s):1\le s\le r\}$ is \emph{noncrossing} \cite{KiwiLoeblMatousek2005,LovaszPlummer}
with respect to $\prec$ if its edges can be indexed such that
\[
i_1\prec\cdots\prec i_r,
\qquad
j_1\prec\cdots\prec j_r.
\]
A directed matching in $D$ is noncrossing with respect to
$\prec$ if its corresponding matching in $\mathcal{B}(D)$
is noncrossing. For either representation $G$, let
$L_\prec(G)$ denote the maximum size of a noncrossing
matching with respect to $\prec$.

We can now relate common subsequences to noncrossing matchings.
Let $V=[n]$. For $\mathbf u,\mathbf v\in\F_q^n$, define
the \emph{relation graph} to be the bipartite
graph
$G_{\mathbf u,\mathbf v}=(V^-\cup V^+,E_{\mathbf u,\mathbf v})$
with
\[
E_{\mathbf u,\mathbf v}
=\{(i,j)\in V\times V:u_i=v_j\},
\] and we denote by $\widetilde G_{\mathbf u,\mathbf v}$ the corresponding loopless
directed graph on $V$ whose edge set is $E_{\mathbf u,\mathbf v}\setminus\{(i,i):i\in V\}$.
For a total order $\prec$ on $[n]$ ($V$, equivalently), let $\mathbf u_\prec$ and $\mathbf v_\prec$ be obtained by
arranging the coordinates of $\mathbf u$ and $\mathbf v$
according to $\prec$.
A noncrossing matching in  $G_{\mathbf u,\mathbf v}$ with edges $(i_1,j_1),\ldots,(i_r,j_r)$
determines the common subsequence
\[
(u_{i_1},\ldots,u_{i_r})=(v_{j_1},\ldots,v_{j_r}).
\]
 Hence we have \begin{equation}\label{lcs ord}
 	\LCS(\mathbf u_\prec,\mathbf v_\prec)
 	=L_\prec(G_{\mathbf u,\mathbf v}).
 \end{equation} The following example illustrates how the vertex order affects 
noncrossing matchings and the value of $L_\prec(G_{\mathbf u,\mathbf v})$.

\begin{example}
 Let $\mathbf u=(0,0,1,0,1)$ and $\mathbf v=(1,1,0,1,0)$ and write the relation graph as $G=G_{\mathbf u,\mathbf v}$. Under the natural order
 $1\prec2\prec3\prec4\prec5$, we have
 $\LCS(\mathbf u,\mathbf v)=L_\prec(G)=3$.
 Under the common order $1\prec^*3\prec^*2\prec^*5\prec^*4$,
 the reordered words are $ \mathbf u_{\prec^*}=(0,1,0,1,0)$ and  $\mathbf v_{\prec^*}=(1,0,1,0,1)$. Their common subsequence $(0,1,0,1)$ is realized by the position pairs
 \[
  M=\{(1,3),(3,2),(2,5),(5,4)\}.
 \]
 Figure~\ref{fig:noncrossing-example} highlights this same matching under both orders. It is crossing under $\prec$ and noncrossing under $\prec^*$. A common subsequence can contain at most two zeros and two ones, so its length is at most $4$ under every coordinate order. Hence
 \[
  L_{\prec^*}(G)=\max_\prec L_\prec(G)=4.
 \]

 % Both panels show the full equal-symbol graph of the two words.
 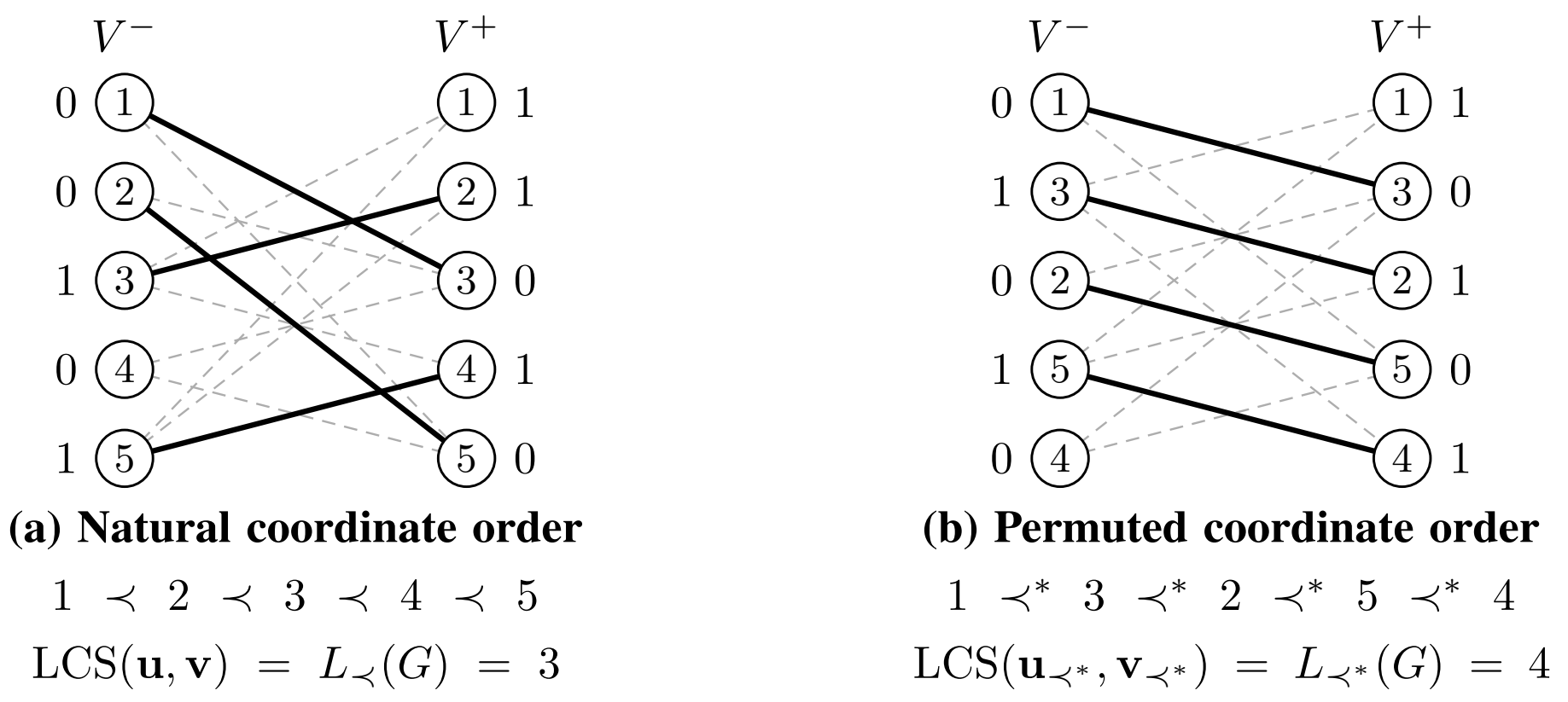
\begin{figure}[H]
  \centering
  \begin{tikzpicture}[x=1cm,y=1cm,font=\small,
   unselected/.style={draw=black!32,line width=0.45pt,densely dashed},
   selected/.style={draw=black,line width=1.2pt},
   vertex/.style={circle,draw=black,fill=white,line width=0.55pt,minimum size=4.4mm,inner sep=0pt,font=\footnotesize}]
   % (a): natural coordinate order on both copies.
   \foreach \v [count=\i from 0] in {1,2,3,4,5}{
    \coordinate (aL\v) at (0,{-0.69*\i});
    \coordinate (aR\v) at (2.65,{-0.69*\i});
   }
   % (b): a common coordinate order with LCS equal to four.
   \foreach \v [count=\i from 0] in {1,3,2,5,4}{
    \coordinate (bL\v) at (7.25,{-0.69*\i});
    \coordinate (bR\v) at (9.90,{-0.69*\i});
   }
   \foreach \p in {a,b}{
    \foreach \u/\v in {1/5,2/3,3/1,3/4,4/3,4/5,5/1,5/2}{
     \draw[unselected] (\p L\u)--(\p R\v);
    }
    \foreach \u/\v in {1/3,3/2,2/5,5/4}{
     \draw[selected] (\p L\u)--(\p R\v);
    }
    \foreach \v/\s in {1/0,2/0,3/1,4/0,5/1}{
     \node[vertex,label=left:{$\s$}] at (\p L\v) {$\v$};
    }
    \foreach \v/\s in {1/1,2/1,3/0,4/1,5/0}{
     \node[vertex,label=right:{$\s$}] at (\p R\v) {$\v$};
    }
   }
   \node[anchor=south] at (0,0.26) {$V^-$};
   \node[anchor=south] at (2.65,0.26) {$V^+$};
   \node[anchor=south] at (7.25,0.26) {$V^-$};
   \node[anchor=south] at (9.90,0.26) {$V^+$};
   \node[align=center,anchor=north,text width=5.6cm] at (1.325,-3.03)
    {\textbf{(a) Natural coordinate order}\\[3pt]
     $1\prec2\prec3\prec4\prec5$\\[3pt]
     $\LCS(\mathbf u,\mathbf v)=L_\prec(G)=3$};
   \node[align=center,anchor=north,text width=6.2cm] at (8.575,-3.03)
    {\textbf{(b) Permuted coordinate order}\\[3pt]
     $1\prec^*3\prec^*2\prec^*5\prec^*4$\\[3pt]
     $\LCS(\mathbf u_{\prec^*},\mathbf v_{\prec^*})=L_{\prec^*}(G)=4$};
  \end{tikzpicture}
  \caption{The relation graph for $\mathbf u=(0,0,1,0,1)$ and $\mathbf v=(1,1,0,1,0)$ under two common coordinate orders. The same  matching $M$ of size $4$ is highlighted in both panels.}
  \label{fig:noncrossing-example}
 \end{figure}
\end{example}

By Lemma \ref{lem:distance-lcs} and Eq.(\ref{lcs ord}), the insdel distance between the two ordered vectors can be written as
\[
d_{\rm I}(\mathbf u_\prec,\mathbf v_\prec)
=2\bigl(n-L_\prec(G_{\mathbf u,\mathbf v})\bigr).
\]
For a code $\cC\subseteq\F_q^n$, write $\cC_\prec:=\{\mathbf u_\prec:\mathbf u\in\cC\}$ for the code obtained by arranging the coordinates of each codeword according to $\prec$. Taking the minimum over distinct codewords gives
\[
d_{\rm I}(\cC_\prec)
=2\left(n-\max_{\mathbf u\ne\mathbf v\in\cC}
L_\prec(G_{\mathbf u,\mathbf v})\right).
\]

Therefore, determining the maximum insdel distance $d_{\rm I}^{\max}(\cC)$ under all coordinate permutations  amounts to minimizing the largest noncrossing matching among the relation graphs of distinct codewords.

The edge count and vertex degrees of a relation graph $G_{\mathbf u,\mathbf v}$  are determined by
symbol multiplicities. For each $a\in\F_q$, define $I_a=\{i\in V:u_i=a\}$ and $J_a=\{j\in V:v_j=a\}$. The edges joining occurrences of $a$ form $I_a\times J_a$. Since the edges of $G_{\mathbf u,\mathbf v}$ are partitioned
into the sets $I_a\times J_a$, we have
\begin{equation}\label{eq:relation-edge-count}
	|E(G_{\mathbf u,\mathbf v})|
	=\sum_{a\in\F_q}|I_a|\,|J_a|.
\end{equation}
Moreover, the degrees of $i^-\in V^-$ and $j^+\in V^+$
are $|J_{u_i}|$ and $|I_{v_j}|$, respectively.
The diagonal edges correspond to coordinates on which $\mathbf u$
and $\mathbf v$ agree, so their number is
$n-d_{\rm H}(\mathbf u,\mathbf v)$. Hence
\[
    \begin{aligned}
    |E(\widetilde G_{\mathbf u,\mathbf v})|
    &=|\{(i,j)\in V\times V:i\ne j,\ u_i=v_j\}|\\
    &=\sum_{a\in\F_q}|I_a|\,|J_a|-n+d_{\rm H}(\mathbf u,\mathbf v).
    \end{aligned}
\]
Removing the diagonal edges from a noncrossing matching in
$G_{\mathbf u,\mathbf v}$ leaves a noncrossing directed matching
in $\widetilde G_{\mathbf u,\mathbf v}$. Consequently,
\begin{equation}\label{eq:diagonal-split}
    \LCS(\mathbf u_\prec,\mathbf v_\prec)
    \le n-d_{\rm H}(\mathbf u,\mathbf v)
    +L_\prec(\widetilde G_{\mathbf u,\mathbf v}).
\end{equation}

For a directed matching $M$ in $\widetilde G_{\mathbf u,\mathbf v}$,
consider the directed subgraph $(V,M)$ with vertex set $V$ and edge set $M$.
Since the edges of $M$ have pairwise distinct tails and pairwise
distinct heads, every vertex of $(V,M)$ has outdegree and
indegree at most one.
More generally, we have the following useful observation
(see \cite{Sagan2020}).
Here, $d_G^+(v)$ and $d_G^-(v)$ denote the outdegree and indegree
of $v$ in a directed graph $G$, respectively. The \emph{weakly connected
components} of $G$ are the connected components obtained by
ignoring edge directions, and  a component is \emph{nontrivial} if
it contains an edge.

\begin{proposition}[\cite{Sagan2020}]\label{prop:path-cycle}
    Let $G$ be a finite loopless directed graph with $d_G^+(v)\le1$ and
    $d_G^-(v)\le1$ for every $v\in V(G)$.
    Then every nontrivial weakly connected component of $G$ is
    either a directed path or a directed cycle.
\end{proposition}

In particular, the nontrivial weakly connected components of
$(V,M)$ are directed paths or directed cycles. In Section~\ref{sec:bounds}, we use this
decomposition to estimate the probability that a fixed directed
matching is noncrossing under a uniformly random vertex order.

\subsection{Algebraic Function Fields and Plane Curves}\label{subsec:curve-prelim}

We recall the facts about function fields and plane curves used in
Section~\ref{sec:explicit}, following \cite{Stichtenoth,Fulton}, and
include short proofs of the auxiliary lemmas needed below.

Let $K$ be an algebraically closed field, and let $\mathcal{X}$ be a
nonsingular projective irreducible curve over $K$, with function field
$F=K(\mathcal{X})$. The places of $F$ correspond to the points of
$\mathcal{X}$. Moreover, $K$-isomorphisms between function fields of
nonsingular projective curves induce isomorphisms of the corresponding
curves \cite[Sec.~7.1, Corollaries~3 and~4]{Fulton}.

Let $\mathbb{P}_F$ denote the set of places of $F$, and let $v_P$
be the normalized discrete valuation at $P\in\mathbb{P}_F$.
A divisor $D=\sum_P n_P P$ is a finite integral sum of places, and
we write $D\ge0$ if all $n_P\ge0$. Since $K$ is algebraically closed,
its degree is $\deg D=\sum_P n_P$. For $z\in F^\times$, its
\emph{principal divisor} and \emph{pole divisor} are defined as
\[
(z)=\sum_{P\in\mathbb{P}_F}v_P(z)P,
\qquad
(z)_\infty=\sum_{\substack{P\in\mathbb{P}_F\\v_P(z)<0}}-v_P(z)P.
\]
For every $z\in F\setminus K$, we have
\cite[Thm.~1.4.11]{Stichtenoth}
\begin{equation}\label{eq:function-degree}
	[F:K(z)]=\deg(z)_\infty.
\end{equation}
The \emph{Riemann--Roch space} associated with a divisor $D$ is
\[
\mathcal{L}(D)=\{z\in F^\times:(z)+D\ge0\}\cup\{0\}.
\]
For an integer $m\ge0$, $\mathcal{L}(mP)$ consists of the functions
having no poles away from $P$ and pole order at most $m$ at $P$,
together with zero. For the level curves $X^r+Y^s=c$ used later,
these spaces have the following description.

\begin{lemma}\label{lem:level-poles-rr}
	Let $r>s\ge2$ be relatively prime integers with
	$\operatorname{char}K\nmid rs$, and let $c\in K^\times$.
	Write $F_c=K(x,y)$ for the function field of $x^r+y^s=c$.
	The functions $x$ and $y$ have the same unique pole $P_\infty$, with
	\[
	(x)_\infty=sP_\infty,
	\qquad
	(y)_\infty=rP_\infty.
	\]
	For every integer $m\ge0$,
	\begin{equation}\label{eq:riemann-roch-basis}
		\mathcal{L}(mP_\infty)
		=\bigl\langle x^iy^j:i\ge0,\ 0\le j<s,\ is+jr\le m\bigr\rangle_K.
	\end{equation}
\end{lemma}

\begin{proof}
	By \cite[Prop.~3.1.15]{Stichtenoth}, the extension $F_c/K(x)$
	has degree $s$ and is totally ramified at infinity, since
	$v_\infty(x^r-c)=-r$ and $\gcd(r,s)=1$.
	Writing $P_\infty$ for the unique place above infinity,
	the relation $y^s=c-x^r$ gives
	$(x)_\infty=sP_\infty$ and $(y)_\infty=rP_\infty$.
	
	Since the affine curve is nonsingular and $y$ is integral over
	$K[x]$, its coordinate ring $K[x,y]$ is the integral closure
	of $K[x]$ in $F_c$, hence the ring of functions regular away
	from $P_\infty$ by \cite[Thm.~3.2.6]{Stichtenoth}.
	Every such function has a unique expansion
	$\sum_{j=0}^{s-1} a_j(x)y^j$ with $a_j(x)\in K[x]$,
	whose monomials $x^iy^j$ have pairwise distinct pole orders
	$is+jr$, and membership in $\mathcal{L}(mP_\infty)$ is therefore
	equivalent to $is+jr\le m$ for every nonzero term, proving
	Eq.~\eqref{eq:riemann-roch-basis}.
\end{proof}

We will also use the following irreducibility statement, which
does not require a restriction on the characteristic or on $c$.

\begin{lemma}\label{lem:level-irreducible}
	Let $r>s\ge2$ be relatively prime integers. Then, for every $c\in K$,
	the polynomial $X^r+Y^s-c$ is irreducible in $K[X,Y]$.
\end{lemma}

\begin{proof}
	View the polynomial as $Y^s+(X^r-c)$ over $K(X)$.
	At the pole of $X$, its constant coefficient has valuation
	$-r$, which is coprime to $s$, and all intermediate
	coefficients are zero.
	Hence \cite[Prop.~3.1.15]{Stichtenoth} gives
	irreducibility in $K(X)[Y]$.
	Gauss's lemma then gives irreducibility in $K[X,Y]$.
\end{proof}

For $A\subseteq K^2$, its \emph{Zariski closure}
$\overline{A}^{\,Z}$ is the common zero set in $K^2$ of all
polynomials in $K[U,V]$ that vanish on $A$
\cite[Sec.~6.1]{Fulton}.

\begin{lemma}\label{lem:plane-image}
	Let $u,v\in F$ be not both constant, and let $S\subseteq\mathcal{X}$
	be a finite set containing all their poles. Set
	\[
	A=\{(u(P),v(P)):P\in\mathcal{X}\setminus S\}.
	\]
	Then the Zariski closure $\Gamma=\overline{A}^{\,Z}$ is an
	irreducible plane curve, and $\Gamma\setminus A$ is finite.
	The substitutions $U\mapsto u$ and $V\mapsto v$ identify
	$K(\Gamma)$ with $K(u,v)\subseteq F$.
\end{lemma}

\begin{proof}
	Let $\rho:K[U,V]\to F$ send $U$ to $u$ and $V$ to $v$.
	A rational function vanishing on $\mathcal{X}\setminus S$
	is zero, so the ideal of polynomials vanishing on $A$ is
	$\ker\rho$. Consequently,
	\[
	K[\Gamma]\cong K[u,v].
	\]
	This is a domain of transcendence degree one over $K$.
	Hence $\Gamma$ is an irreducible plane curve and
	$K(\Gamma)=K(u,v)$ under the indicated identification.
	
	Let $\overline{\Gamma}\subseteq\mathbb{P}^2$ be the projective
	closure of $\Gamma$.
	The rational map $P\mapsto[1:u(P):v(P)]$ extends to a morphism
	$\Phi:\mathcal{X}\to\overline{\Gamma}$ because $\mathcal{X}$
	is nonsingular \cite[Sec.~7.1, Corollary~1]{Fulton}.
	This morphism is nonconstant and therefore surjective.
	It follows that $\Gamma\setminus A\subseteq\Phi(S)$,
	which is finite.
\end{proof}

Finally, the following affine form of B\'ezout's theorem
follows from the projective version by taking projective
closures \cite[Sec.~5.3, Corollary~3]{Fulton}.

\begin{theorem}\label{lem:bezout}
	Two plane curves over $K$ of degrees $d_1$ and $d_2$ with no common
	irreducible component have at most $d_1d_2$ distinct common affine points.
\end{theorem}

\section{Bounds on the Insdel Distance of Linear Codes under Coordinate Permutations}\label{sec:bounds}
We establish lower and upper bounds on the insdel distance attainable under coordinate permutations. We first derive tail estimates for noncrossing directed matchings under a common random vertex order. Applying these estimates to relation graphs of codeword pairs gives high-probability lower bounds in terms of the Hamming distance and symbol multiplicities, with applications to MDS codes, full-length RS codes, and two non-MDS evaluation codes. We then establish an upper bound for projective linear codes that holds for every coordinate order and, together with the probabilistic lower bounds, determines the optimal asymptotic distance under a symbol-multiplicity condition.

\subsection{Maximum Noncrossing Matchings of Relation Graphs under a Common Random Order}

Since both vertex classes of a relation graph share the same
coordinate order, we first estimate the probability that a fixed
directed matching is noncrossing under a uniformly random
vertex order.

\begin{lemma}\label{lem:fixed-matching}
Let $M$ be a fixed directed matching of size $r\ge1$ in a loopless
directed graph $G=(V,E)$, and consider the directed subgraph $(V,M)$.
If $(V,M)$ contains a directed cycle, then no total order of $V$
makes $M$ noncrossing. Otherwise, its nontrivial weakly connected
components are directed paths. If $(V,M)$ has $c$ nontrivial weakly connected components, then
a uniformly random total order of $V$ satisfies
\[
 \mathbf{P}\left(M\text{ is noncrossing}\right)
 \le\frac{1}{c!(r-c)!}
 \le\frac{2^r}{r!}.
\]
\end{lemma}

\begin{proof}

By Proposition~\ref{prop:path-cycle}, every nontrivial weakly
connected component of $(V,M)$ is a directed path or a directed
cycle. Suppose that $M$ is noncrossing with respect to a vertex
order $\prec$ and contains a directed cycle $C$. Let $V_0$ be the
vertex set of $C$. The edges of $C$ define a permutation
$\sigma:V_0\to V_0$, where $(v,\sigma(v))$ is the edge leaving $v$.
Since $M$ is noncrossing, $u\prec v$ implies
$\sigma(u)\prec\sigma(v)$ for all $u,v\in V_0$, but
every increasing permutation of a finite totally ordered set
is the identity. Hence $\sigma(v)=v$ for every $v\in V_0$,
contradicting the assumption that $G$ is loopless.

Suppose now that the nontrivial weakly connected components of
$(V,M)$ are $c$ directed paths. Since these paths have $r$ edges
in total, the support of $M$ has $r+c$ vertices. Identify the
positions in the order induced on this support with $[r+c]$,
and let $D$ and $R$ be the sets of positions occupied by tails
and heads, respectively. The complements $D^c$ and $R^c$
consist of the terminal and initial vertices, respectively.
They are disjoint sets of size $c$, so there are at most
\[
 \binom{r+c}{c}\binom rc
\]
possible ordered pairs $(D^c,R^c)$.

Write $D=\{d_1<\cdots<d_r\}$ and $R=\{s_1<\cdots<s_r\}$.
A noncrossing matching must join $d_i$ to $s_i$ for every $i$.
Thus $D$ and $R$ determine the edges on the ordered support.
The $c$ path components of $(V,M)$ can be assigned to the
resulting path components on $[r+c]$ in at most $c!$ ways.
Each assignment determines all vertex positions along the paths.
Since the relative order of the support is uniform,
\[
 \mathbf{P}\left(M\text{ is noncrossing}\right)
 \le\frac{\binom{r+c}{c}\binom rc\,c!}{(r+c)!}
 =\frac{1}{c!(r-c)!}
 =\frac{\binom rc}{r!}
 \le\frac{2^r}{r!}.
\]
\end{proof}

We next show that the probability that a random order admits a large noncrossing directed matching can be bounded in terms of the edge count, with a sharper estimate available when the maximum indegree and outdegree are also controlled.

\begin{proposition}\label{prop:matching-tail}
Let $G=(V,E)$ be a loopless directed graph with at most $m$ edges, where $m\ge1$ is an integer. Choose a total order $\prec$ of $V$ uniformly at random. For every integer $1\le r\le m$,
\begin{equation}\label{eq:coarse-tail}
 \mathbf{P}\left(L_\prec(G)\ge r\right)\le\binom mr\frac{2^r}{r!}.
\end{equation}
If every indegree and outdegree is at most $\Delta$, then
\begin{equation}\label{eq:refined-tail}
 \mathbf{P}\left(L_\prec(G)\ge r\right)
 \le\frac{m^r}{(r!)^2}
 \exp\!\left(2\sqrt{\frac{\Delta r^3}{m}}\right).
\end{equation}
\end{proposition}

\begin{proof}
The event $L_\prec(G)\ge r$ requires an $r$-edge directed matching to be noncrossing. There are at most $\binom mr$ such matchings, so Lemma~\ref{lem:fixed-matching} gives Eq.~\eqref{eq:coarse-tail} by a union bound.

For the second estimate, Lemma~\ref{lem:fixed-matching} allows us
to restrict attention to directed matchings $M$ for which every
nontrivial weakly connected component of $(V,M)$ is a directed
path. Fix the number $c$ of these path components and write
$a_v=d_G^+(v)$. The number of choices for their initial edges is
at most
\[
 \sum_{\substack{S\subseteq V\\|S|=c}}\prod_{v\in S}a_v
 \le\frac{1}{c!}\left(\sum_v a_v\right)^c
 \le\frac{m^c}{c!}.
\]

Label the chosen initial edges using a fixed auxiliary order of their tails. The path lengths form a vector of positive integers with sum $r$,  and there are $\binom{r-1}{c-1}$ possible length vectors. Once these data are fixed, each remaining edge extends a path in at most $\Delta$ ways. Allowing repeated vertices can only increase the count, so there are at most
\[
 \frac{m^c}{c!}\binom{r-1}{c-1}\Delta^{r-c}
\]
directed matchings $M$ with $r$ edges and $c$ nontrivial path
components in $(V,M)$. Lemma~\ref{lem:fixed-matching} now gives
\[
 \mathbf{P}\left(L_\prec(G)\ge r\right)
 \le\sum_{c=1}^r
 \frac{m^c\Delta^{r-c}}{(c!)^2(r-c)!}
 \binom{r-1}{c-1}.
\]

Put $j=r-c$ and factor out $m^r/(r!)^2$. The remaining factor in the $j$-th term is
\[
 \frac{1}{(j!)^2}
 \left(\frac{\Delta}{m}\right)^j
 \left(\frac{r!}{(r-j)!}\right)^2
 \frac{(r-1)!}{(r-1-j)!}
 \le\frac{1}{(j!)^2}
 \left(\frac{\Delta r^3}{m}\right)^j.
\]
For $x\ge0$, we have
\[
 \sum_{j\ge0}\frac{x^j}{(j!)^2}
 \le\left(\sum_{j\ge0}\frac{x^{j/2}}{j!}\right)^2
 =e^{2\sqrt x}.
\]
Taking $x=\Delta r^3/m$ proves Eq.~\eqref{eq:refined-tail}.
\end{proof}

\subsection{Insdel Distance of Linear Codes under Random Coordinate Permutations}
For a linear code $\cC$, let $\cC^+=\cC+\langle\one\rangle$.
The following identity for the maximum symbol multiplicity of nonconstant codewords is immediate.
\begin{lemma}\label{lem:symbol-multiplicity}
For a linear code $\cC$ containing a nonconstant word,
\[
	\max_{\substack{
			\mathbf u\in\cC\setminus\langle\one\rangle\\
			a\in\F_q
	}}
	|\{i:u_i=a\}|
	=n-d_{\rm H}(\cC^+).
\]

\end{lemma}

In what follows, we write $\mu(\cC):=n-d_{\rm H}(\cC^+)$ for the maximum symbol multiplicity of nonconstant codewords in $\cC$. Combining Lemma~\ref{lem:symbol-multiplicity} and Proposition \ref {prop:matching-tail}, we  derive the following high-probability lower bound on the insdel distance.

\begin{theorem}\label{thm:linear-random}
Let $\cC$ be a sequence of $[n,k]_q$ linear codes containing a nonconstant word. Suppose that $\mu(\cC)=o(n)$ and $k\log q=o(\sqrt{n\mu(\cC)})$ as $n\to\infty$. For every fixed $\eps>0$ and all sufficiently large $n$, a uniformly random coordinate order satisfies
\[
 \mathbf{P}\left(d_{\rm I}(\cC_\prec)\ge2d_{\rm H}(\cC)-(2e+\eps)\sqrt{n\mu(\cC)}\right)
 \ge1-\exp\!\left(-\frac\eps2\sqrt{n\mu(\cC)}\right)
 \longrightarrow1.
\]
\end{theorem}

\begin{proof}
We first obtain finite estimates for an arbitrary $[n,k]_q$ linear code containing a nonconstant word. For distinct nonconstant codewords, Lemma~\ref{lem:symbol-multiplicity} and Eq.~\eqref{eq:relation-edge-count} in Section \ref{sec:prelim} give $ |E(\widetilde G_{\mathbf u,\mathbf v})|\le n\mu(\cC)$ and $d^+_{\widetilde G_{\mathbf u,\mathbf v}}(i),
\ d^-_{\widetilde G_{\mathbf u,\mathbf v}}(i)\le \mu(\cC)$.

For a constant codeword $a\one$ and a nonconstant codeword
$\mathbf u$, Lemma~\ref{lem:symbol-multiplicity} gives
\[
\LCS(a\one,\mathbf u_\prec)
=|\{i:u_i=a\}|\le \mu(\cC).
\]
Hence, any pair
with LCS greater than $\mu(\cC)$ must consist of two nonconstant
codewords.

For a pair of distinct nonconstant codewords, the relation graph has
$n-d_{\rm H}(\mathbf u,\mathbf v)\le n-d_{\rm H}(\cC)$ diagonal edges.
By Eq.~\eqref{eq:diagonal-split} in Section \ref{sec:prelim}, a common subsequence of length
greater than $n-d_{\rm H}(\cC)+r$ gives a noncrossing directed matching
with at least $r$ edges in $\widetilde G_{\mathbf u,\mathbf v}$.
Applying Proposition~\ref{prop:matching-tail} with $m=n\mu(\cC)$ and
$\Delta=\mu(\cC)$, and taking a union bound over fewer than $q^{2k}$
ordered pairs of distinct nonconstant codewords gives, for every
integer $1\le r\le n$,
\[
 \mathbf{P}\left(\LCS(\cC_\prec)>\max\{\mu(\cC),n-d_{\rm H}(\cC)+r\}\right)
 \le q^{2k}\binom{n\mu(\cC)}{r}\frac{2^r}{r!}
 \le q^{2k}\left(\frac{2e^2n\mu(\cC)}{r^2}\right)^r
\]
and
\begin{equation}\label{eq:linear-refined-tail}
 \mathbf{P}\left(\LCS(\cC_\prec)>\max\{\mu(\cC),n-d_{\rm H}(\cC)+r\}\right)
 \le q^{2k}\frac{(n\mu(\cC))^r}{(r!)^2}
 \exp\!\left(2\sqrt{\frac{r^3}{n}}\right).
\end{equation}

Now assume $\mu(\cC)=o(n)$ and $k\log q=o(\sqrt{n\mu(\cC)})$. Fix $\eps>0$ and put $ r=\left\lfloor\left(e+\frac\eps2\right)\sqrt{n\mu(\cC)}\right\rfloor$. For sufficiently large $n$, we have $1\le r\le n$ and $\mu(\cC)\le r$. Since $2r\le(2e+\eps)\sqrt{n\mu(\cC)}$, Eq.~\eqref{eq:linear-refined-tail} gives
\begin{align*}
 &\mathbf{P}\left(d_{\rm I}(\cC_\prec)<2d_{\rm H}(\cC)-(2e+\eps)\sqrt{n\mu(\cC)}\right)\\
 &\qquad\le\exp\!\left(
 2k\log q-2r\log\frac{r}{e\sqrt{n\mu(\cC)}}
 +2r\sqrt{\frac rn}\right).
\end{align*}
The inequality $x\log x\ge x-1$ for $x>0$ implies
\[
 r\log\frac{r}{e\sqrt{n\mu(\cC)}}\ge r-e\sqrt{n\mu(\cC)}.
\]
Hence the exponent is at most
\[
 -\eps\sqrt{n\mu(\cC)}+2+2k\log q+2r\sqrt{\frac rn}.
\]
The last three terms are $o(\sqrt{n\mu(\cC)})$. For sufficiently large $n$, the failure probability is therefore at most $\exp(-\eps\sqrt{n\mu(\cC)}/2)$.
\end{proof}

When the symbol multiplicities are at most a fixed multiple of their average $n/q$, the dimension condition follows from the Singleton bound. The resulting distance lower bound has the form $2n-O(n/\sqrt q)$.

\begin{corollary}\label{cor:linear-square-root}
Let $\cC$ be a sequence of $[n,k]_q$ linear codes containing a nonconstant word, where $q\to\infty$. Suppose that
\[
 n-d_{\rm H}(\cC^+)\le K\frac nq
\]
for a fixed constant $K>0$. For every fixed $\eps>0$ and all sufficiently large $q$, a uniformly random coordinate order satisfies
\[
 \mathbf{P}\left(d_{\rm I}(\cC_\prec)\ge2n-(2e+\eps)\sqrt K\,\frac n{\sqrt q}\right)
 \ge1-\exp\!\left(-\frac{\eps n}{4\sqrt q}\right)
 \longrightarrow1.
\]
\end{corollary}

\begin{proof}
Lemma~\ref{lem:symbol-multiplicity} and the Singleton bound give
\[
 \frac nq\le \mu(\cC)\le K\frac nq,
 \qquad
 k\le\dim\cC^+\le \mu(\cC)+1\le2\mu(\cC).
\]
In particular, $\mu(\cC)\ge1$ and $n\ge q/K$. Thus $\mu(\cC)=o(n)$ and
\[
 \frac{k\log q}{\sqrt{n\mu(\cC)}}
 \le2\sqrt K\,\frac{\log q}{\sqrt q}
 \longrightarrow0.
\]
Apply Theorem~\ref{thm:linear-random} with $\eps/2$. Since $d_{\rm H}(\cC)\ge n-\mu(\cC)$, its distance lower bound is at least
\[
 2n-2\mu(\cC)-\left(2e+\frac\eps2\right)\sqrt{n\mu(\cC)}
 \ge2n-(2e+\eps)\sqrt K\,\frac n{\sqrt q}
\]
for sufficiently large $q$. Here $\sqrt{n\mu(\cC)}\le\sqrt K\,n/\sqrt q$ and $2\mu(\cC)=o(n/\sqrt q)$. Also, $\sqrt{n\mu(\cC)}\ge n/\sqrt q$; thus the failure probability is at most $\exp(-\eps n/(4\sqrt q))$. Finally, $n\ge q/K$ implies $n/\sqrt q\to\infty$.
\end{proof}

We first apply Theorem~\ref{thm:linear-random} to MDS codes
containing the all-one vector.

\begin{proposition}\label{prop:mds-random}
Let $\cC$ be a sequence of $[n,k]_q$ MDS codes containing $\one$,
where $2\le k<n$ and $(k-1)(\log q)^2=o(n)$ as $n\to\infty$.
For every fixed $\eps>0$ and all sufficiently large $n$,
a uniformly random coordinate order $\prec$ satisfies
\[
\begin{aligned}
 &\mathbf{P}\left(d_{\rm I}(\cC_\prec)
 \ge 2(n-k+1)-(2e+\eps)\sqrt{(k-1)n}\right)\\
 &\qquad\ge
 1-\exp\!\left(-\frac{\eps}{2}\sqrt{(k-1)n}\right)
 \longrightarrow1.
\end{aligned}
\]
\end{proposition}

\begin{proof}
Since $\one\in\cC$, we have $\cC^+=\cC$ and
$\mu(\cC)=n-d_{\rm H}(\cC)=k-1$.
The result follows from Theorem~\ref{thm:linear-random}
because $(k-1)(\log q)^2=o(n)$ implies
$\mu(\cC)=o(n)$ and $k\log q=o(\sqrt{n\mu(\cC)})$.
\end{proof}

Specializing Proposition~\ref{prop:mds-random} to full-length
RS codes gives the following corollary.

\begin{corollary}\label{cor:rs-random}
Let $k\ge2$ satisfy $k=o(q/(\log q)^2)$ as $q\to\infty$,
and let $\boldsymbol\alpha$ be a uniformly random ordering of $\F_q$.
For every fixed $\eps>0$ and all sufficiently large $q$,
\[
\begin{aligned}
 &\mathbf{P}\left(
 d_{\rm I}(\RS_{q,k}(\boldsymbol\alpha))
 \ge 2(q-k+1)-(2e+\eps)\sqrt{(k-1)q}\right)\\
 &\qquad\ge
 1-\exp\!\left(-\frac{\eps}{2}\sqrt{(k-1)q}\right)
 \longrightarrow1.
\end{aligned}
\]
\end{corollary}

\begin{proof}
Since full-length RS codes are MDS codes containing $\one$,
the result follows from Proposition~\ref{prop:mds-random}
with $n=q$.
\end{proof}

\begin{remark}\label{rem:rs-beelen}
Let $\cC=\RS_{q,2}(\boldsymbol\alpha)$, where $\boldsymbol\alpha$
is a uniformly random ordering of $\F_q$.
Replacing the count $\binom{r+c}{r}^2$ in
\cite[proof of Prop.~2]{BeelenEtAl} by the exact count
$\binom{r+c}{r}\binom rc$ for pairs of increasing index
sequences of length $r$ with a fixed union of size $r+c$
gives, for $3\le r<q$,
\[
 \mathbf{P}\bigl(\LCS(\cC)\ge r\bigr)
 \le q^2\sum_{c=1}^{\min\{r,q-r\}}
 \frac{q^c}{(r-c)!(c!)^2}
 \le q^2\frac{q^r}{(r!)^2}
 \exp\!\left(\frac{r^2}{q}\right).
\]
The last inequality follows by setting $j=r-c$ and using
$r!/(r-j)!\le r^j$.
For every fixed $\eps>0$, taking
$r=\lceil(e+\eps/2)\sqrt q\rceil$ and using $r!\ge(r/e)^r$
yields $d_{\rm I}(\cC)\ge2q-(2e+\eps)\sqrt q$ with probability
at least $1-\exp(-\Omega_\eps(\sqrt q))$ for all sufficiently large $q$.
\end{remark}

We next apply Corollary~\ref{cor:linear-square-root}
to two non-MDS codes obtained by evaluating affine functions.

\begin{example}\label{ex:rm-affine}
Let $q\ge3$ be a prime power. The code
\[
 \cC=\operatorname{RM}_q(1,2)
 =\{(a+bx+cy)_{(x,y)\in\F_q^2}:a,b,c\in\F_q\}
\]
is the first-order Reed--Muller code on $\F_q^2$
\cite{DelsarteGoethalsMacWilliams1970}.
Since every nonconstant affine function takes each value exactly
$q$ times, $\cC$ has parameters $[q^2,3,q^2-q]_q$ and is not MDS.
It contains $\one$ and satisfies
\[
 \mu(\cC)=n-d_{\rm H}(\cC^+)=q=\frac nq.
\]
For every fixed $\eps>0$ and all sufficiently large $q$,
Corollary~\ref{cor:linear-square-root} with $K=1$ gives
\[
\begin{aligned}
 &\mathbf{P}\left(
 d_{\rm I}(\cC_\prec)
 \ge 2q^2-(2e+\eps)q^{3/2}\right)\\
 &\qquad\ge
 1-\exp\!\left(-\frac{\eps}{4}q^{3/2}\right)
 \longrightarrow1
\end{aligned}
\]
for a uniformly random coordinate order $\prec$.
\end{example}

A second example is obtained by evaluating affine functions
on the affine rational points of a Hermitian curve.

\begin{example}\label{ex:hermitian-three}
Let $q$ be a prime power and let $S$ be the set of affine
$\F_{q^2}$-rational points of the Hermitian curve
\[
 y^q+y=x^{q+1}.
\]
The Hermitian curve code
\[
 \cC=\{(a+bx+cy)_{(x,y)\in S}:a,b,c\in\F_{q^2}\}
\]
has parameters $[q^3,3,q^3-q-1]_{q^2}$ and is not MDS
\cite[Sec.~8.3]{Stichtenoth}.
Since $\one\in\cC$, we have
\[
 \mu(\cC)=n-d_{\rm H}(\cC^+)=q+1\le2\frac n{q^2},
 \qquad n=q^3.
\]
For every fixed $\eps>0$ and all sufficiently large $q$,
Corollary~\ref{cor:linear-square-root} with alphabet size $q^2$
and $K=2$ gives
\[
\begin{aligned}
 &\mathbf{P}\left(
 d_{\rm I}(\cC_\prec)
 \ge 2q^3-(2e+\eps)\sqrt2\,q^2\right)\\
 &\qquad\ge
 1-\exp\!\left(-\frac{\eps}{4}q^2\right)
 \longrightarrow1
\end{aligned}
\]
for a uniformly random coordinate order $\prec$.
\end{example}

\subsection{An Upper Bound on the Insdel Distance of Projective Linear Codes}
\label{subsec:optimal}

We now complement the preceding probabilistic lower bounds with an alphabet-dependent upper bound on the insdel distance of projective linear codes that holds for every coordinate order. Combined with the high-probability lower bound of Corollary~\ref{cor:linear-square-root}, it determines the maximum attainable insdel distance to be $2n-\Theta(n/\sqrt q)$ when the nonconstant codewords have symbol multiplicities $O(n/q)$.

A linear code is projective if any two columns of a generator
matrix are linearly independent. This ensures pairwise independence
of coordinate matching indicators for two independent random codewords.
A second-moment argument over consecutive blocks then yields
a long common subsequence.
\begin{theorem}\label{thm:projective-upper}
	Let $\cC$ be a projective $[n,k]_q$ linear code with $k\ge2$.
	Then
	\begin{equation}\label{eq:projective-simple}
		d_{\rm I}^{\max}(\cC)
		\le 2n-\left(1-\frac1{\lfloor\sqrt q\rfloor}\right)
		\left\lfloor\frac{n}{\lfloor\sqrt q\rfloor}\right\rfloor.
	\end{equation}
\end{theorem}
\begin{proof}
	Fix a coordinate order, and let
	$g_1,\ldots,g_n\in\F_q^k$ be the generator columns.
	Set $b=\lfloor\sqrt q\rfloor$ and $M=\lfloor n/b\rfloor$,
	and take $M$ disjoint consecutive blocks of length $b$.
	Choose $U,V\in\F_q^k$ independently and uniformly.
	For each block $B$, let
	\[
	X_B=\sum_{i,j\in B}
	\mathbf1_{\{U\cdot g_i=V\cdot g_j\}}.
	\]
	Since the generator columns are nonzero and pairwise
	nonproportional, distinct equalities in this sum impose
	independent linear conditions on $(U,V)$.
	Thus the indicators are pairwise independent, each with
	mean $1/q$, and
	\[
	\mathbb E[X_B]=\frac{b^2}{q},
	\qquad
	\mathbb E[X_B^2]
	=\frac{b^2}{q}+\frac{b^2(b^2-1)}{q^2}.
	\]
	By Cauchy--Schwarz,
	\[
		\mathbf P(X_B>0)
		\ge\frac{\mathbb E[X_B]^2}{\mathbb E[X_B^2]}
		=\frac{b^2}{q+b^2-1}.
	\]
	
	Let $Y$ count the blocks with $X_B>0$. Then
	\[
	\mathbb E[Y]\ge M\frac{b^2}{q+b^2-1}.
	\]
	Since $Y=M$ whenever $U=V$, and
	$\mathbf P(U=V)=q^{-k}$,
	\[
	\mathbb E[Y\mid U\ne V]
	\ge M\frac{b^2/(q+b^2-1)-q^{-k}}{1-q^{-k}}.
	\]
	As $q\le b^2+2b$ and $q^k\ge(b+1)^2$, we have
	\[
	\frac{b^2/(q+b^2-1)-q^{-k}}{1-q^{-k}}
	\ge\frac{b/(2(b+1))-q^{-k}}{1-q^{-k}}
	\ge\frac{b-1}{2b}.
	\]
	Consequently,
	\[
	\mathbb E[Y\mid U\ne V]\ge\frac{b-1}{2b}M.
	\]
	
	Hence there exist $U\ne V$ for which
	$Y\ge(b-1)M/(2b)$.
	The corresponding codewords are distinct, and one
	matching pair from each of these blocks gives a common
	subsequence of length at least $Y$. Therefore,
	\[
	\begin{aligned}
		d_{\rm I}(\cC)
		&\le 2n-2Y\\
		&\le 2n-\left(1-\frac1b\right)M\\
		&=2n-\left(1-\frac1{\lfloor\sqrt q\rfloor}\right)
		\left\lfloor\frac{n}{\lfloor\sqrt q\rfloor}\right\rfloor.
	\end{aligned}
	\]
\end{proof}

\begin{remark}\label{rem:upper-comparison}
    We compare Theorem~\ref{thm:projective-upper} with earlier
    upper bounds for projective $[n,k]_q$ codes with $k\ge2$ as
    $q\to\infty$, assuming
    $\mu(\cC)=n-d_{\rm H}(\cC^+)=O(n/q)$.
    Table~\ref{tab:upper-comparison} states the upper bounds directly
    for the insdel distance of an arbitrarily ordered code.
    The earlier bounds in the table have right-hand sides of the form
    $2n-O(n/q)$, whereas Theorem~\ref{thm:projective-upper} gives
    $2n-\Theta(n/\sqrt q)$.
    Together with Corollary~\ref{cor:linear-square-root}, this yields
    $d_{\rm I}^{\max}(\cC)=2n-\Theta(n/\sqrt q)$,
    a uniformly random coordinate order attains the same asymptotic
    distance with high probability.

   	\begin{table*}[h]
   	\centering
   	\caption{Upper bounds on $d_{\rm I}(\cC_\prec)$ for an arbitrary
   		coordinate order $\prec$ of a projective $[n,k]_q$ code with
   		$\mu(\cC)=O(n/q)$.}
   	\label{tab:upper-comparison}
   	\small
   	\renewcommand{\arraystretch}{1.4}
   	\setlength{\tabcolsep}{8pt}
   	\begin{tabular}{@{}lcc@{}}
   		\hline
   		Bounds & Upper bounds on $d_{\rm I}(\cC_\prec)$ & Asymptotic form of the bounds\\
   		\hline
   		Half-Singleton \cite{ChengGuruswamiHaeuplerLi}
   		& $2n-(4k-4)$ & $2n-O(n/q)$\\
   		Strict half-Singleton \cite{JiZhengChenWang2023}
   		& $2n-(4k-2)$ & $2n-O(n/q)$\\
   		Direct bound
   		& $2d_{\rm H}(\cC)$ & $2n-O(n/q)$\\
   		Strict direct bound \cite{JiZhengChenWang2023}
   		& $2(d_{\rm H}(\cC)-1)$ & $2n-O(n/q)$\\
   		Chen \cite[Thm.~2.1]{Chen2022}
   		& $2n-2(h_\prec+z_\prec-1)$ & $2n-O(n/q)$\\
   		Xie et al.\ \cite[Thm.~2.1]{XieChenQuLiu2024}
   		& $2n-2\lceil L/(q-1)\rceil$ & $2n-O(n/q)$\\
   		\hline
   		Theorem~\ref{thm:projective-upper}
   		& $\displaystyle 2n-
   		\left(1-\frac1{\lfloor\sqrt q\rfloor}\right)
   		\left\lfloor\frac n{\lfloor\sqrt q\rfloor}\right\rfloor$
   		& $2n-\Theta(n/\sqrt q)$\\
   		\hline
   	\end{tabular}
   \end{table*}
   
   The strict half-Singleton bound applies only if
   $\one\notin\cC$ \cite{JiZhengChenWang2023}.
   For the strict direct bound, Corollary~6.2 of
   \cite{JiZhengChenWang2023} applies here for sufficiently large
   $q$: projectivity holds by assumption, while
   $d_{\rm H}(\cC)\ge n-\mu(\cC)>(n+1)/2$.
   The Singleton bound and $\cC\subseteq\cC^+$ give
   $1\le k-1\le n-d_{\rm H}(\cC)\le\mu(\cC)=O(n/q)$,
   and thus $n/q=\Omega(1)$.
   In Chen's bound \cite[Thm.~2.1]{Chen2022}, $h_\prec$ is the
   number of selected linearly independent generator columns, and
   $z_\prec$ is the number of zero coordinates of a chosen nonzero
   codeword before the first or after the last selected position
   in the order $\prec$.
   Hence $h_\prec\le k$ and $z_\prec\le n-d_{\rm H}(\cC)$ for every
   $\prec$, giving $h_\prec+z_\prec-1=O(n/q)$ uniformly in $\prec$.
   In Xie et al.'s bound \cite{XieChenQuLiu2024}, $L$ is a common
   lower bound on the Hamming weights of two linearly independent
   codewords. Since $L\le n$, we have
   $\lceil L/(q-1)\rceil=O(n/q)$.
\end{remark}

\section{Explicit Orderings of Full-Length Reed--Solomon Codes with Large Insdel Distance}\label{sec:explicit}

We now return to full-length RS codes and seek explicit orderings with distance $2q-\Theta(\sqrt q)$ for fixed dimension. Corollary~\ref{cor:rs-random} and Theorem~\ref{thm:projective-upper} show that random orderings attain this asymptotic distance. Over square fields, we construct such orderings by partitioning the evaluation points into consecutive blocks and using algebraic intersection bounds to control common values between blocks.

\begin{lemma}\label{lem:partition}
	Let $G$ be a bipartite graph with vertex classes $V^-$ and $V^+$,
	both copies of a finite set $V$.
	Let $V=V_1\sqcup\cdots\sqcup V_m$ be a partition of $V$ into nonempty
	subsets, and fix an order $\prec$ on $V$ in which these subsets
	appear as consecutive blocks in the order $V_1,\ldots,V_m$.
	The order within each block is arbitrary, and the same order
	is used on both vertex classes. For each $i$, let $V_i^-$ and $V_i^+$ denote the copies of
	$V_i$ in $V^-$ and $V^+$, respectively, and write
	$G[V_i^-,V_j^+]$ for the subgraph induced by
	$V_i^-\cup V_j^+$.
	Suppose that $G[V_i^-,V_j^+]$ has ordinary matching number
	at most $b$ for every $1\le i,j\le m$. Then
	\[
	L_\prec(G)\le(2m-1)b.
	\]
\end{lemma}

\begin{proof}
	Let $M$ be a nonempty noncrossing matching, and list its edges
	in increasing order of their left endpoints. Their right
	endpoints occur in the same order. Thus, the corresponding
	block indices are nondecreasing in both coordinates.
	After removing consecutive repetitions, let
	$(i_1,j_1),\ldots,(i_t,j_t)$ be the block pairs visited.
	Since at least one index increases at each step,
	\[
	t-1\le(i_t-i_1)+(j_t-j_1)\le2(m-1).
	\]
	The edges of $M$ belonging to any one block pair form an
	ordinary matching of size at most $b$. Consequently,
	$|M|\le tb\le(2m-1)b$.
\end{proof}

In the proof of Lemma~\ref{lem:partition}, the block indices
$(i,j)$ along a noncrossing matching are nondecreasing in
both coordinates. The following figure shows a
possible sequence of five distinct block pairs for $m=3$,
attaining the maximum $5$.

\begin{figure}[htbp]
	\centering
	\begin{tikzpicture}[x=1cm,y=0.85cm,font=\small]
		% Visited block pairs
		\foreach \i/\j/\t in {
			1/1/1,1/2/2,2/2/3,2/3/4,3/3/5
		} {
			\fill[black!10]
			({\j-1},{-\i}) rectangle (\j,{1-\i});
			\node at ({\j-0.5},{0.5-\i}) {\t};
		}
		
		% Grid
		\draw[black!55,line width=0.4pt]
		(0,-3) rectangle (3,0);
		\foreach \i in {1,2} {
			\draw[black!55,line width=0.4pt]
			(\i,-3) -- (\i,0);
			\draw[black!55,line width=0.4pt]
			(0,-\i) -- (3,-\i);
		}
		
		% Block labels
		\foreach \i in {1,2,3} {
			\node[above=4pt] at ({\i-0.5},0) {$V_\i^+$};
			\node[left=5pt] at (0,{0.5-\i}) {$V_\i^-$};
		}
	\end{tikzpicture}
\caption{Five successive block pairs for $m=3$.}
	\label{fig:partition-path}
\end{figure}

For general dimension $k$, we give the following construction of full-length RS codes. For each fixed $k$, the resulting code has distance $2q-\Theta(\sqrt q)$, hence the construction is order-optimal.

\begin{theorem}\label{thm:explicit-general}
Let $k\ge2$, and let $q=q_0^2$ be a square prime power of characteristic $p>k+1$. Choose $\theta\in\F_q\setminus\F_{q_0}$. For each $c\in\F_{q_0}$, define
\[
 S_c=\{x+\theta y:x,y\in\F_{q_0},\ x^{k+1}+y^k=c\}.
\]
List the nonempty sets $S_c$ consecutively to obtain an ordering $\boldsymbol\alpha$ of $\F_q$. Both the order of the sets and the orders within them may be arbitrary. Then
\begin{equation}\label{eq:explicit-general-distance}
 d_{\rm I}(\RS_{q,k}(\boldsymbol\alpha))
 \ge2q-4\bigl(1+(k+1)^2(k-1)^3\bigr)\sqrt q.
\end{equation}

For fixed $k$, such an ordering can be constructed deterministically
in time polynomial in $q$.

\end{theorem}

To apply Lemma~\ref{lem:partition}, we bound matchings between nonzero blocks through intersections of plane curves. Work over $K=\overline{\F}_q$, and let
$\sigma:a\mapsto a^{q_0}$ be the Frobenius automorphism of
$\F_q/\F_{q_0}$. For $f\in\F_q[T]$, write $f^\sigma$ for
the polynomial obtained by applying $\sigma$ coefficientwise to $f$.  For $c\in\F_{q_0}^*$, let $F_c$ be the function field in
Lemma~\ref{lem:level-poles-rr} with $(r,s)=(k+1,k)$, and let
$\mathcal{X}_c$ be its nonsingular projective model.
We identify $\mathcal{X}_c\setminus\{P_\infty\}$ with the affine
curve $X^{k+1}+Y^k=c$.
For a nonconstant polynomial $f\in\F_q[T]$, let $\Gamma_{f,c}$
be the Zariski closure in $K^2$ of the image of
$\mathcal{X}_c\setminus\{P_\infty\}$ under the map
\[
(x,y)\longmapsto
\bigl(f(x+\theta y),f^\sigma(x+\theta^{q_0}y)\bigr).
\]
The first coordinate function has pole order $(k+1)\deg f$ at $P_\infty$. Hence, Lemma~\ref{lem:plane-image} with $S=\{P_\infty\}$ shows that $\Gamma_{f,c}$ is an irreducible plane curve.

\begin{lemma}\label{lem:image-curves}
Let $c,d\in\F_{q_0}^*$ and let $f,h\in\F_q[T]$ be nonconstant polynomials of degree less than $k$. Then
\begin{equation}\label{eq:image-degree}
 \deg\Gamma_{f,c}\le(k+1)\deg f.
\end{equation}
Moreover, $\Gamma_{f,c}\ne\Gamma_{h,d}$ if $f\ne h$.
\end{lemma}

\begin{proof}
Let $m=\deg f$, by Lemma~\ref{lem:plane-image}, a general
affine line $\ell$ meets $\Gamma_{f,c}$ in
$\deg\Gamma_{f,c}$ distinct points, all lying in the
image of the map. Substituting the coordinate functions
of the map into the equation of $\ell$ gives a plane
curve of degree at most $m$ that intersects
$X^{k+1}+Y^k=c$ in at least $\deg\Gamma_{f,c}$
distinct points. B\'ezout's theorem
(Theorem~\ref{lem:bezout}) now yields
\eqref{eq:image-degree}.

By Lemma~\ref{lem:plane-image}, we identify
\[
 K(\Gamma_{f,c})
 =K\bigl(f(x+\theta y),f^\sigma(x+\theta^{q_0}y)\bigr)
 \subseteq F_c.
\]
Let $a_m$ be the leading coefficient of $f$, and let $N=m(k+1)$.
Both generators have pole divisor $NP_\infty$. Consider the normalized difference
\[
 w=
 \frac{f(x+\theta y)}{a_m\theta^m}
 -
 \frac{f^\sigma(x+\theta^{q_0}y)}{a_m^{q_0}\theta^{mq_0}}
 \in K(\Gamma_{f,c}).
\]
The $y^m$ terms cancel, while the coefficient of $xy^{m-1}$ is
$m(\theta^{-1}-\theta^{-q_0})\ne0$, since $m<p$ and
$\theta\notin\F_{q_0}$. A monomial $x^iy^j$ has pole order
$ki+(k+1)j$. Thus $xy^{m-1}$ has pole order $N-1$,
whereas the other degree $m$ terms have pole order at most $N-2$
and the lower-degree terms have pole order at most $(m-1)(k+1)$.
Since $w$ has no finite poles, $(w)_\infty=(N-1)P_\infty$.
Eq.~\eqref{eq:function-degree} therefore gives
\[
 [F_c:K(f(x+\theta y))]=N,
 \qquad
 [F_c:K(w)]=N-1.
\]
Both subfields lie in $K(\Gamma_{f,c})$, so the tower law shows that
$[F_c:K(\Gamma_{f,c})]$ divides both $N$ and $N-1$.
These integers are coprime, and hence $K(\Gamma_{f,c})=F_c$.

Suppose $\Gamma_{f,c}=\Gamma_{h,d}$. The preceding calculation identifies $F_c$ and $F_d$ with the function field of their common image. Let $x',y'\in F_c$ be the images of the generators of $F_d$. The first coordinate of the image has a unique pole on each curve.
Thus their places at infinity correspond, and comparison of its pole orders gives $\deg h=m$.
By Lemma~\ref{lem:level-poles-rr},
\[
 \mathcal{L}(kP_\infty)=\langle1,x\rangle_K,
 \qquad
 \mathcal{L}((k+1)P_\infty)=\langle1,x,y\rangle_K.
\]
Since $x'$ and $y'$ have pole orders exactly $k$ and $k+1$, respectively, they take the form
\[
 x'=ax+b,\qquad y'=ux+vy+e,\qquad av\ne0.
\]
Equality of the coordinate functions on the common image yields
\[
 \begin{aligned}
 f(x+\theta y)&=h(x'+\theta y'),\\
 f^\sigma(x+\theta^{q_0}y)&=h^\sigma(x'+\theta^{q_0}y').
 \end{aligned}
\]
After substituting these expressions and replacing $x,y$ by $X,Y$, the differences of the two sides are polynomials of degree at most $m<k+1$ that vanish on $X^{k+1}+Y^k=c$. The ideal of this irreducible curve is generated by $X^{k+1}+Y^k-c$. Each difference is therefore divisible by this polynomial, and the degree bound forces both differences to be zero polynomials. Comparing the $X^m$ and $X^{m-1}Y$ coefficients gives
\[
 a+\theta u=v=a+\theta^{q_0}u.
\]
Thus $u=0$ and $v=a$. The relation $(x')^{k+1}+(y')^k=d$ now shows that $(aX+b)^{k+1}+(aY+e)^k-d$ is a scalar multiple of $X^{k+1}+Y^k-c$. Comparing the coefficients of $X^{k+1}$ and $Y^k$ gives $a^{k+1}=a^k$, and hence $a=1$. Since $p>k+1$, the coefficients of $X^k$ and $Y^{k-1}$ give $b=e=0$. It follows that $c=d$, and the first coordinate identity reduces to $f(x+\theta y)=h(x+\theta y)$. Since $x+\theta y$ is nonconstant, this implies $f=h$.
\end{proof}

Now we can prove Theorem~\ref{thm:explicit-general}.

\begin{proof}[Proof of Theorem~\ref{thm:explicit-general}]
Let $f,h\in\F_q[T]$ be distinct nonconstant polynomials of degree less than $k$. The edges between blocks $S_c$ and $S_d$ in their relation graph correspond to pairs of evaluation points with the same polynomial value. For $x+\theta y\in S_c$ with $x,y\in\F_{q_0}$, the corresponding point of $\Gamma_{f,c}$ is $(f(x+\theta y),f(x+\theta y)^{q_0})$. Hence, when $c,d\ne0$, distinct common values in $f(S_c)\cap h(S_d)$ give distinct points of $\Gamma_{f,c}\cap\Gamma_{h,d}$. The curves are distinct by Lemma~\ref{lem:image-curves}, so B\'ezout's theorem bounds the number of common values by $(k+1)^2\deg f\,\deg h$. Each value contributes at most $\min\{\deg f,\deg h\}$ edges to a matching, since a nonconstant polynomial takes any given value at most as many times as its degree. Thus the ordinary matching number between any two nonzero blocks is at most
\[
 (k+1)^2\deg f\,\deg h\min\{\deg f,\deg h\}
 \le(k+1)^2(k-1)^3.
\]

We claim that $|S_0|=q_0$. Apart from $(0,0)$, the solutions of
$x^{k+1}+y^k=0$ over $\F_{q_0}$ are parametrized by
\[
(x,y)=(-t^k,-t^{k+1}),\qquad t\in\F_{q_0}^{\times},
\]
where $t$ is uniquely recovered as $y/x$, giving exactly
$q_0-1$ nonzero solutions. Deleting all edges incident to $S_0$ on either side of a
noncrossing matching therefore removes at most $2q_0$ edges,
leaving a matching supported on at most $q_0-1$ consecutive
blocks on each side, with the preceding matching bound
applicable to every pair of remaining blocks. Lemma~\ref{lem:partition} gives
\[
 \begin{aligned}
 \LCS(f(\boldsymbol\alpha),h(\boldsymbol\alpha))
 &\le 2q_0+(2q_0-3)(k+1)^2(k-1)^3\\
 &\le 2\bigl(1+(k+1)^2(k-1)^3\bigr)q_0.
 \end{aligned}
\]
The LCS is at most $k-1$ if exactly one polynomial is constant,
and zero for distinct constants. This completes the proof.
\end{proof}

In dimension two, we show that alternative choices of the level set
polynomials yield the same asymptotic distance bound
in characteristics $2$ and $3$.

\begin{proposition}\label{prop:explicit-two}
Let $q=q_0^2$ have characteristic $p\in\{2,3\}$. Choose $\theta\in\F_q\setminus\F_{q_0}$, and set
\[
 (r,s)=
 \begin{cases}
 (5,2),&p=3,\\
 (5,3),&p=2.
 \end{cases}
\]
For each $c\in\F_{q_0}$, define
\[
 S_c=\{x+\theta y:x,y\in\F_{q_0},\ x^r+y^s=c\}.
\]
List the nonempty sets $S_c$ consecutively to obtain an ordering $\boldsymbol\alpha$ of $\F_q$. Both the order of the sets and the orders within them may be arbitrary. Then
\begin{equation}\label{eq:explicit-two-distance}
 d_{\rm I}(\RS_{q,2}(\boldsymbol\alpha))
 \ge2q-100\sqrt q+50.
\end{equation}

\end{proposition}

\begin{proof}
The exponents satisfy $\gcd(r,s)=1$, $p\nmid rs$, and $r-s=p$. Write $ \theta^2=\tau\theta+\nu$, where $\tau,\nu\in\F_{q_0} $ and $\nu\ne0$. For every nonidentity affine map $\gamma(z)=az+b$ with $a\ne0$, we claim that
\begin{equation}\label{eq:affine-level-intersection}
 |S_c\cap\gamma^{-1}(S_d)|\le25.
\end{equation}
Write $a=a_0+a_1\theta$ and $b=b_0+b_1\theta$. The points counted in Eq.~\eqref{eq:affine-level-intersection} correspond to common affine points of
\[
 X^r+Y^s=c
\]
and
\[
 (a_0X+\nu a_1Y+b_0)^r+
 \bigl(a_1X+(a_0+\tau a_1)Y+b_1\bigr)^s=d.
\]
Both curves are absolutely irreducible of degree $r=5$ by Lemma~\ref{lem:level-irreducible} and the invertibility of multiplication by $a$.

If the curves had a common component, their defining polynomials would differ by a nonzero scalar factor $\lambda$. Comparing homogeneous terms of degree $r$ would give
\[
 (a_0X+\nu a_1Y)^r=\lambda X^r.
\]
Thus $a_1=0$ and $\lambda=a_0^r$. The coefficients of $X^{r-1}$ and $Y^{s-1}$ then give $b_0=b_1=0$, since $p\nmid rs$. Comparing the coefficients of $Y^s$ gives $a_0^{r-s}=1$. As $r-s=p$, this forces $a_0=1$, contradicting $\gamma\ne\mathrm{id}$. Then B\'ezout's theorem proves Eq.~\eqref{eq:affine-level-intersection}.

For distinct nonconstant affine polynomials $f,h$, the relation $f(x)=h(y)$ is $y=\gamma(x)$ for the nonidentity affine map $\gamma=h^{-1}\circ f$. Its relation graph is a matching with at most $25$ edges in each block pair. Lemma~\ref{lem:partition} gives
\[
 \LCS(f(\boldsymbol\alpha),h(\boldsymbol\alpha))
 \le25(2q_0-1).
\]
A distinct pair involving a constant polynomial has LCS at most one. This completes the proof.
\end{proof}

Together with Theorem~\ref{thm:explicit-general} for $k=2$, this proposition gives deterministic orderings with distance $2q-\Theta(\sqrt q)$ over every square field.

\section{Conclusion}\label{sec:conclusion}

In this paper, we investigated how coordinate orderings affected the insertion-deletion performance of linear codes over large alphabets. The connection between common subsequences and noncrossing matchings provided a common perspective on the distance attainable by random permutations and the design of explicit orderings. For projective linear codes with suitably controlled symbol multiplicities, our results showed that random orderings attained the best possible asymptotic order $\Theta(n/\sqrt q) $ of the gap from $2n$. This understanding also led to explicit orderings for full-length Reed--Solomon codes of fixed dimension over square fields, which achieved a gap of the same square-root order.

A more precise understanding of the maximum attainable distance remains of interest, particularly the constants in the gap of order $n/\sqrt q$ from $2n$. In particular, for full-length Reed--Solomon codes, it would be interesting to determine the coefficient of the $\sqrt q$ term for fixed dimensions and to understand how the maximum distance changes when the dimension grows with the field size. Extending the explicit constructions to arbitrary finite fields would also be a natural direction for further study.

\end{document}